\documentclass[aps,prd,twocolumn,groupedaddress,nofootinbib]{revtex4-1}

\usepackage{dsfont}
\usepackage{natbib}
\usepackage{amssymb}
\usepackage{graphicx}
\usepackage{amsmath}
\usepackage{url}
\usepackage{bm}
\usepackage{epstopdf}
\usepackage{color}
\usepackage{enumitem}
\usepackage{mathrsfs}

\newcommand{\di}{\partial}
\newcommand{\f}{\mathscr{F}}

\def\be {\begin{equation}}

\def\ee  {\end{equation}}

\def\bea {\begin{eqnarray}}

\def\eea {\end{eqnarray}}

\def\nn {\nonumber}

\begin{document}

\title{Infrared modification of gravity from conformal symmetry}

\author{Jack Gegenberg}
\email{geg@unb.ca}
\affiliation{Department of Mathematics and Statistics, University of New Brunswick, Fredericton, NB, Canada E3B 5A3}

\author{Shohreh Rahmati}
\email{shohreh.rahmati@unb.ca}
\affiliation{Department of Mathematics and Statistics, University of New Brunswick, Fredericton, NB, Canada E3B 5A3}

\author{Sanjeev S.\ Seahra}
\email{sseahra@unb.ca}
\affiliation{Department of Mathematics and Statistics, University of New Brunswick, Fredericton, NB, Canada E3B 5A3}

\date{\today}

\begin{abstract}

We reconsider a gauge theory of gravity in which the gauge group is the conformal group SO(4,2) and the action is of the Yang-Mills form, quadratic in the curvature.  The resulting gravitational theory exhibits local conformal symmetry and reduces to Weyl-squared gravity under certain conditions.  When the theory is linearized about flat spacetime, we find that matter which couples to the generators of special conformal transformations reproduces Newton's inverse square law.  Conversely, matter which couples to generators of translations induces a constant and possibly repulsive force far from the source, which may be relevant for explaining the late time acceleration of the universe.  The coupling constant of theory is dimensionless, which means that it is potentially renormalizable.

\end{abstract}

\maketitle

\section{Introduction}

General relativity (GR) is a well-tested theory of gravity on scales ranging from microns to the size of the solar system.  But on larger scales, cracks appear in the edifice of observational support for the theory.  The gravitational interaction over galactic distances requires dark matter to account for rotation curves and dynamics of clusters, while over cosmological distances we require dark energy to drive accelerated expansion.  While it is quite possible, and some may argue probable, that one of both of these effects have an explanation within particle physics, it seems worthwhile to explore the possibility that GR itself is not the correct theory of gravity on large scales.

An intriguing line of attack involves replacing GR with a theory respecting local conformal (Weyl) symmetry.  There are diverse motivations for such an approach (see \cite{Hooft:2014daa} for a recent discussion):  One of the earliest involves noting that Maxwell's equations are not only invariant under the Poincare group of transformations ISO(3,1), they are also invariant under the larger SO(4,2) group of conformal transformations of Minkowski spacetime \cite{cunningham,*bateman} (as are other important massless field theories \cite{dirac,*drew}, including the standard model with zero Higgs mass).  It seems logical that the theory of gravity should have the same symmetries as electromagnetism, at least locally, yet GR retains local ISO(3,1), not SO(4,2), symmetry.

Shortly after the appearance of GR, Weyl \cite{weyl} and Bach \cite{bach} tried to rectify this by writing down a locally conformally invariant theory of gravity and electromagnetism.  This approach failed as a unified theory, giving unacceptable gravity-EM couplings, and also failed to reproduce solar system dynamics as linear gravity was governed by a fourth order Poisson equation.  To recover the inverse square law in conformal gravity, Mannheim has recently suggested that point sources are described by highly singular distributions involving derivatives of $\delta$-functions \cite{Mannheim:2007ug,Mannheim:1992tr,*Mannheim:1999bu,*Mannheim:2011ds}.

Another approach to alternative theories of gravity with a long history is based on Yang-Mills (YM) gauge theories. The YM action (here written assuming a compact gauge group),
\be
S= - \frac{1}{2 g_\text{YM}^{2}}\int d^4x \sqrt{-g}g^{\mu\nu}g^{\alpha\beta}\text{Tr}( \textbf{F}_{\mu\alpha} \textbf{F}_{\nu\beta}),
\ee
is quadratic in the curvature $\mathbf{F}_{\alpha\beta}$ of a Lie-algebra valued connection $\textbf{A}_{\alpha}$ on a principle bundle over the spacetime manifold $(M,g)$, where $g_{\alpha\beta}$ is a {\it given} non-dynamical Lorentzian metric on $M$.  The coupling constant $g_\text{YM}$ is dimensionless, which is what makes the theory perturbatively renormalizable.  By contrast, the Einstein-Hilbert action of GR is linear in the curvature of the Christoffel connection, there is no background (the metric is dynamical), and the coupling constant has dimension $(\text{mass})^{2}$.  The latter presents one of the main obstructions to quantizing GR.

To construct a YM gauge theory of gravity one first chooses a gauge group, and then identifies components of the gauge potential $\mathbf{A}_{\alpha}$ with a vierbein $e^{a}_{\alpha}$ (or soldering form) and spin-connection $\omega^{ab}_{\alpha}$ on $(M,g)$.  This introduces an explicit algebraic dependence of the metric on the gauge potential, implying that YM gravity models generally exhibit less gauge symmetry than their purely YM cousins.  Also, like the Einstein-Cartan action for GR, the YM gravity action is in first order form, but unlike the Einstein-Cartan action, the torsion of the connection is not constrained to be zero \cite{Utiyama:1956sy, *kibble,*townsend,*macman,*PhysRevLett.42.1021,*hayshir,*ivnied,Wheeler:1991ff,*Hazboun:2013lra,*Wheeler:2013ora,Huang:2008ik,*Huang:2009nc,*Huang:2009sk,*Lu:2013sai,*Lu:2013bt}.

If one selects one of the Poincare, de Sitter, or anti-de Sitter groups as the gauge group, the identification of  $e^{a}_{\alpha}$ and spin-connection $\omega^{ab}_{\alpha}$ exhausts all the components of $\mathbf{A}_{\alpha}$.  (The de Sitter case has been investigated in \cite{Huang:2008ik,*Huang:2009nc,*Huang:2009sk,*Lu:2013sai,*Lu:2013bt}.)  Here, we follow Wheeler et al  \cite{Wheeler:1991ff,*Hazboun:2013lra,*Wheeler:2013ora} and consider the gauging of the SO(4,2) group.  In this case, there are extra components of the gauge potential associated with special conformal transformations and dilatations.  This allows for more general matter gravity coupling than in GR.  The resulting equations exhibit local conformal symmetry.

By analyzing the linearization of this model about flat space, we find that matter that couples directly to the tetrad $e^{a}_{\alpha}$ sources a long range gravitational potential that may be repulsive.  Conversely, matter which couples to the special conformal transformations sources a Newton-like gravitational potential that gives rise to the familiar inverse square law.  Hence, in this model it is in principle possible to realize infrared modifications of gravity while still maintaining Newtonian gravity in the solar system.

\section{The model}

Our model is based on the SO(4,2) conformal group of Minkowski spacetime, which is the largest group of transformations that leaves null geodesics invariant.  With $a,b=0\ldots3$ and $A,B=1\ldots 15$, the fifteen generators $\mathbf{J}_{A}$ of this group can be subdivided into: four translations $\mathbf{P}_{a}$, four special conformal transformations $\mathbf{K}_{a}$, six Lorentz rotations $\mathbf{J}_{ab}=-\mathbf{J}_{ba}$, and one dilatation $\mathbf{D}$.  The non-zero commutators of generators are:
\begin{align}
\nn [{\bf J}_{ab}, {\bf J}_{cd}] & =
\eta_{ad}{\bf J}_{bc} +
\eta_{bc}{\bf J}_{ad} +
\eta_{ac}{\bf J}_{db} +
\eta_{bd}{\bf J}_{ca}, \\ \nn
[{\bf P}_a, {\bf J}_{bc}] & =
\eta_{ba}{\bf P}_c - \eta_{ca}{\bf P}_b, \quad
[{\bf D}, {\bf K}_{a}] = {\bf K}_{a},  \\ \nn
[{\bf K}_a, {\bf J}_{bc}]  & =
\eta_{ba}{\bf K}_c - \eta_{ca}{\bf K}_b,  \quad
[{\bf P}_a, {\bf D}] = {\bf P}_a, \\
 [{\bf P}_a, {\bf K}_{b}] & =
2 (\eta_{ab}{\bf D} - {\bf J}_{ab}),
\end{align}
where $\eta_{ab}$ is the Minkowski metric.

We define an so(4,2)-Lie algebra-valued vector potential by
\begin{equation}
\mathbf{A}_{\alpha} = A^{A}_{\alpha} \mathbf{J}_{A} =   e^{a}_{\alpha} \mathbf{P}_{a} + l^{a}_{\alpha} \mathbf{K}_{a} + \omega_{\alpha}^{ab} \mathbf{J}_{ab} + q_{\alpha} \mathbf{D},
\end{equation}
where $\alpha = 0 \ldots 3$ is a spacetime index.  The components of associated field strength $\mathbf{F}_{\alpha\beta} = F^{A}_{\alpha\beta} \mathbf{J}_{A}$ are given by
\begin{equation}
	F^{A}_{\alpha\beta} = \di_{\alpha} A^{A}_{\beta} - \di_{\beta} A^{A}_{\alpha} + f^{A}{}_{BC} A^{B}_{\alpha} A^{C}_{\beta},
\end{equation}
with the structure constants defined by $[\mathbf{J}_{A},\mathbf{J}_{B} ] = f^{C}{}_{AB} \mathbf{J}_{C}$.

We identify various components of $\mathbf{A}_{\alpha}$ in the $\mathbf{J}_{A}$ basis with geometric quantities in a 4-dimensional Lorentzian manifold $M$ with metric $g_{\alpha\beta}$ and affine connection $\Gamma^{\alpha}{}_{\beta\delta}$.  In particular, we take $e^{a}_{\alpha}$ as the components of an  orthonormal frame fields on $M$, with $\omega^{ab}_{\alpha}$ as the associated connection one-forms.  Hence, the metric and connection are given by:
\begin{equation}\label{eq:tetrad postulate}
	g_{\alpha\beta} = \eta_{ab} e^{a}_{\alpha} e^{b}_{\beta}, \quad \Gamma^{\gamma}{}_{\alpha\beta} = e^{\gamma}_{a}(\di_{\alpha} e_{\beta}^{a}  + \omega^{ac}_{\alpha} e_{c\beta}).
\end{equation}
In these expressions, lowercase Greek and Latin indices are raised and lowered with $g_{\alpha\beta}$ and $\eta_{ab}$, respectively.  The curvature one-forms are anti-symmetric in their frame indices $\omega^{(ab)}_{\alpha} = 0$, from which it follows that the affine connection is metric compatible \cite{Carroll:2004st}:
\begin{equation}
	0 = \nabla_{\alpha} g_{\beta\gamma},
\end{equation}
where $\nabla_{\alpha}$ is the derivative operator defined by $\Gamma^{\alpha}{}_{\beta\delta}$.  The Riemann curvature and torsion tensors of $M$ are given by:
\begin{align}
	R^{\mu\nu}{}_{\alpha\beta} & = e^{\mu}_{a} e^{\nu}_{b} (d\omega^{ab} + \omega^{ac} \wedge \omega_{c}{}^{b})_{\alpha\beta}, \\
	T^{\alpha}{}_{\beta\gamma} & = e^{\alpha}_{a} (de^{a} + \omega^{ac}\wedge e_{c})_{\beta\gamma}.\label{eq:torsion def}
\end{align}
Note that we do not assume $T^{\alpha}{}_{\beta\gamma} = 2\Gamma^{\alpha}{}_{[\beta\gamma]}=0$.

To write down the action of our model, we start with the curved spacetime Yang-Mills action assuming a compact gauge group:
\begin{align}
	S & = - \frac{1}{2g^{2}_\text{YM}} \int d^{4}x \sqrt{-g} g^{\alpha\mu} g^{\beta\nu} \text{Tr} \,( \mathbf{F}_{\alpha\beta} \mathbf{F}_{\mu\nu}) + S_\text{m}.
\end{align}
Here, $S_\text{m} = S_\text{m}[A_{\alpha}^{A},\psi]$ is the action for matter $\psi$.  To get this into a form suitable for a non compact gauge group such as SO(4,2), we merely substitute in $\mathbf{F}_{\alpha\beta} = F^{A}_{\alpha\beta} \mathbf{J}_{A}$ :
\begin{align}
	S = - \frac{1}{2g^{2}_\text{YM}} \int d^{4}x \sqrt{-g} g^{\alpha\mu} g^{\beta\nu} h_{AB} F^{A}_{\alpha\beta} F^{B}_{\mu\nu} + S_\text{m},\label{eq:action}
\end{align}
where $h_{AB} = \text{Tr}(\mathbf{J}_{A} \mathbf{J}_{B})$ is the Cartan-Killing metric on so(4,2).  This can be related to the structure constants via the formula:
\begin{equation}\label{eq:killing metric}
	h_{AB} = f^{M}{}_{AN} f^{N}{}_{BM}.
\end{equation}
The non-trivial components of $h_{AB}$ are:
\begin{gather}
\nn h_{a\bar{b}} = h_{\bar{a}b}=-2\eta_{ab}, \quad h_{14,14} = 2.\\
h_{[ab][cd]} = h_{[cd][ab]}=-4\eta_{a[c}\eta_{d]b}.
\end{gather}
The notation here is that $a,\bar{a}=0,1,2,3$ denote components in the direction of translations $\mathbf{P}_a$ and special conformal transformations $\mathbf{K}_a$, respectively.  The six indices $[ab]$ consist of $[12],[23],[31],[01],[02],[03]$ and denote directions along  the distinct non-zero generators $\mathbf{J}_{ab}$ of Lorentz transformations.  Finally, the index $14$ denotes the component in the direction of the generator $\mathbf{D}$ of dilatations.

We view (\ref{eq:action}) and (\ref{eq:killing metric}) as the defining relationships for our model .  Notice that the generators $\mathbf{J}_{A}$ do not explicitly appear in either of these formulae, yet the particular choice of basis for so(4,2) does influence $h_{AB}$ and $F^{A}_{\alpha\beta}$.  That is, if we change basis according to
\begin{equation}
	\tilde{\mathbf{J}}_{A} = C_{A}{}^{B} \mathbf{J}_{B},
\end{equation}
where $C_{A}{}^{B}$ is a matrix such that $C_{A}{}^{D} C_{D}{}^{B} = \delta_{A}{}^{B}$, then the field strength components and Killing metric transform as
\begin{equation}
	\tilde{h}_{AB} = C_{A}{}^{C} C_{B}{}^{D} h_{CD}, \quad \tilde{F}^{A}_{\alpha\beta} = C_{B}{}^{A} F^{B}_{\alpha\beta}.
\end{equation}
It follows that (\ref{eq:action}) itself is invariant under such a transformation; i.e., the action for our model is invariant under a change of so(4,2) basis.  Furthermore, the action is manifestly diffeomorphism invariant.

Making use of the definitions above, we can rewrite the action (\ref{eq:action}) exclusively in terms of spacetime tensors:
\begin{multline}\label{eq:spacetime action}
	S = \frac{2}{g_\text{YM}^{2}} \int d^{4}x \sqrt{-g} \Big[ (R_{\alpha\beta\gamma\delta} - \tfrac{1}{2} \phi_{\alpha\beta\gamma\delta})^{2} \\ + 2  (\nabla^{\alpha}f^{\mu\beta}+ f^{\mu\sigma}T_{\sigma}{}^{\alpha\beta}-f^{\mu\alpha}q^{\beta}) \\ \times  (T_{\mu\alpha\beta} +2 g_{\mu[\alpha} q_{\beta]} ) + \\ (\di_{[\alpha} q_{\beta]} + \f_{\alpha\beta})^{2}  \Big] + S_{m}.
\end{multline}
Here, we have defined:
\begin{gather}
	\nn f_{\alpha\beta} := \eta_{ab} e_{\alpha}^{a} l_{\beta}^{b}, \quad \f_{\alpha\beta} := \tfrac{1}{2} f_{[\alpha\beta]}, \\
	\phi^{\alpha\beta\gamma\delta}  := g^{\gamma[\alpha} f^{\beta]\delta}  - g^{\delta[\alpha} f^{\beta]\gamma}. \label{eq:phi def}
\end{gather}
We demonstrate in Appendix \ref{sec:weyl squared} that under certain restrictive circumstances, (\ref{eq:spacetime action}) reduces to the action of Weyl-squared gravity.

Variation of the action (\ref{eq:action}) with respect to the gauge potential yields the equation of motion:
\begin{equation}\label{eq:EOM 1}
	 D_{\mu} F^{B\mu\nu} = k^{B\nu} + j^{B\nu},
\end{equation}
where $D_{\mu}$ is the gauge covariant derivative:
\begin{equation}
	D_{\mu} F^{B\mu\nu} := \hat\nabla_{\mu}  F^{B\mu\nu} + f^{B}{}_{CD} A^{C}_{\mu} F^{D\mu\nu},
\end{equation}
and $\hat\nabla_{\mu}$ is the derivative operator defined from the Levi-Civita connection:
\begin{equation}
	\hat{\Gamma}^{\alpha}{}_{\beta\gamma} = \tfrac{1}{2} g^{\alpha\rho} (\di_{\beta} g_{\rho\gamma} + \di_{\gamma} g_{\rho\beta} - \di_{\rho} g_{\beta\gamma}).
\end{equation}
Note that $\nabla_{\alpha} = \hat\nabla_{\alpha}$ if and only if $T^{\alpha}{}_{\beta\gamma}=0$.  The currents in (\ref{eq:EOM 1}) are given by:
\begin{equation}
	k^{A\nu} \mathbf{J}_{A} = \frac{\tau^{\mu\nu} e^{b}_{\mu} \mathbf{K}_{b}}{2}, \quad j^{\nu}_{C} = - \frac{g_\text{YM}^{2}}{2\sqrt{-g}} \frac{\delta(\sqrt{-g} \mathcal{L}_\text{m})}{\delta A^{C}_{\nu}},
\end{equation}
where $\mathcal{L}_\text{m}$ is the Lagrangian density of the matter fields, $j_{C}^{\nu} = h_{BC} j^{B\nu}$ and
\begin{equation}
	\tau^{\rho\sigma} = h_{AB}\left( F^{A\rho\mu} F^{B\sigma}{}_{\mu} - \tfrac{1}{4} g^{\rho\sigma} F^{A}_{\mu\nu} F^{B\mu\nu} \right).
\end{equation}
We parametrize the matter current as
\begin{equation}
j^{B\nu} \mathbf{J}_{B} =  a^{a\nu} \mathbf{P}_{a} + b^{a\nu} \mathbf{K}_{a} + c^{ab\nu} \mathbf{J}_{ab} + d^{\nu} \mathbf{D},
\end{equation}
and define $a^{\alpha\nu} := e^{\alpha}_{a} a^{a\nu}$, etc.  The structure of $h_{AB}$ implies that $a_{\alpha\beta}$ is proportional to the functional derivative of the matter action with respect to $l^{a}_{\alpha}$; that is, $a_{\alpha\beta}$ characterizes matter which couples to the generators of special conformal transformations.  Similarly, $b_{\alpha\beta}$, $c_{\alpha\beta\gamma}$, and $d_{\alpha}$ describe matter with coupling to $e^{a}_{\alpha}$, $\omega^{ab}_{\alpha}$, or $q_{\alpha}$, respectively.

\section{Gauge symmetries}

The action (\ref{eq:action}) is \emph{not} invariant under the usual infinitesimal YM gauge transformations,
\begin{equation}
	A^{A}_{\alpha} \mapsto A^{A}_{\alpha} +  \partial_{\alpha}\epsilon^{A}+ f^{A}{}_{BC} A^{B}_{\alpha} \epsilon^{C},
\end{equation}
with arbitrary gauge parameters $\epsilon^{A}$.  The reason is that although (\ref{eq:action}) closely resembles the familiar YM action, it differs in one key respect: The metric carries an explicit dependence on the gauge potential via the identification $g_{\alpha\beta} = \eta_{ab} e^{a}_{\alpha} e^{b}_{\beta}$, which breaks the full YM gauge symmetry.

To see this explicitly, let us parametrize an arbitrary gauge transformation as
\begin{equation}
	\epsilon^{A} \mathbf{J}_{A} =  \chi^{a} \mathbf{P}_{a} + \lambda^{a} \mathbf{K}_{a} + \Lambda^{ab} \mathbf{J}_{ab} + \Omega \mathbf{D}.
\end{equation}
We note following gauge transformations:
\begin{subequations}
\begin{align}
	\delta e^{d}_{\alpha} & = \di_{\alpha} \chi^{d} - e^{c}_{\alpha}\Lambda^{d}{}_{c} + \tfrac{1}{2} e^{d}_{\alpha} \Omega + \omega^{dc}_{\alpha} \chi_{c} - \tfrac{1}{2} q_{\alpha} \chi^{d},\\
	\delta l^{d}_{\alpha} & = \di_{\alpha} \lambda^{d}  - l^{c} _{\alpha} \Lambda^{d}{}_{c} - \tfrac{1}{2} l^{d}_{\alpha} \Omega + \omega^{dc}_{\alpha} \lambda_{c} + \tfrac{1}{2} q_{\alpha} \lambda^{d}, \\
	\nn \delta\omega_{\alpha}^{ab} & = \di_{\alpha} \Lambda^{ab} +  \omega_{\alpha}{}^{a}{}_{c} \Lambda^{cb} + \omega_{\alpha}{}^{b}{}_{c} \Lambda^{ac}  \\ & + \tfrac{1}{2} (\lambda^{[a} e^{b]}_{\alpha} - \chi^{[a} l^{b]}_{\alpha}), \\
	\delta q_{\alpha}& = \di_{\alpha} \Omega + \tfrac{1}{2} (\lambda_{\alpha} -  \chi_{\beta} f^{\beta}{}_{\alpha}), \\
	\delta g_{\alpha\beta} & = 2 \nabla_{(\alpha} \chi_{\beta)} + \Omega g_{\alpha\beta} - \tfrac{1}{2} (q_{\alpha} \chi_{\beta} + q_{\beta} \chi_{\alpha}), \\
	\delta f_{\alpha\beta} & = (\nabla_{\beta} + \tfrac{1}{2} q_{\beta}) \lambda_{\alpha} + f^{\mu}{}_{\beta}(\nabla_{\alpha} - \tfrac{1}{2} q_{\alpha}) \chi_{\mu},
\end{align}
\end{subequations}
where we have written $\chi_{\sigma} = e_{\sigma a}\chi^{a}$, etc.  Temporarily switching off the matter fields, we find the variation of the action under this gauge transformation is
\begin{equation}\label{eq:gauge variation of S}
	\delta S = -\frac{2}{g^{2}_\text{YM}} \int d^{4}x \sqrt{-g} \chi_{\sigma}(\nabla_{\rho} \tau^{\rho\sigma}+ \tfrac{1}{2} q_{\rho} \tau^{\rho\sigma}).
\end{equation}
Hence, the action is invariant only if we take $\chi_{\sigma} = 0$.  In other words, this theory is invariant under an eleven parameter subgroup of SO(4,2) gauge transformations parametrized by
\begin{equation}\label{eq:gauge 1}
	\epsilon^{A} \mathbf{J}_{A} =\lambda^{a} \mathbf{K}_{a} + \Lambda^{ab} \mathbf{J}_{ab} + \Omega \mathbf{D}.
\end{equation}
The absence of the $\mathbf{P}_{a}$ generators in (\ref{eq:gauge 1}) is not surprising: as in GR, the Poincar\'{e} translational symmetry of the Minkowski metric is supplanted by the diffeomorphism invariance of curved space in our model.\footnote{When thinking about gauge transformations in this model, one may be tempted to interpret $\mathbf{P}_{a}$ as the generator or \emph{diffeomorphisms} rather than \emph{translations}.  As is clear from (\ref{eq:action}) and (\ref{eq:gauge variation of S}), the action is invariant under diffeomorphisms but not invariant under gauge transformations generated by $\mathbf{P}_{a}$.  Therefore, $\mathbf{P}_{a}$ is not the generator of diffeomorphims.}

Under the transformation (\ref{eq:gauge 1}), the component of $\mathbf{A}_{\alpha}$ parallel to $\mathbf{D}$ transforms as
\begin{equation}
	\delta q_{\alpha} = \di_{\alpha}\Omega + \tfrac{1}{2} \lambda_{\alpha}.
\end{equation}
It is obvious that via a simple series of gauge transformations of the form $\epsilon^{A} \mathbf{J}_{A} = \lambda^{a} \mathbf{K}_{a}$ we can impose the gauge condition $q_{\alpha}=0$.  For the remainder of this paper, we will work exclusively in such a gauge.  Note that there is still some residual gauge freedom: the condition $q_{\alpha}=0$ is preserved under transformations parametrized by:
\begin{equation}\label{eq:gauge 2}
	\epsilon^{A} \mathbf{J}_{A} = -2 e^{a\alpha} \di_{\alpha} \Omega \, \mathbf{K}_{a} + \Lambda^{ab} \mathbf{J}_{ab} + \Omega \mathbf{D}.
\end{equation}
Under this, the frame fields and metric transform as
\begin{equation}\label{eq:gauge 3}
	\delta e^{a}_{\alpha} = (-\Lambda^{a}{}_{b} + \tfrac{1}{2} \Omega \delta^{a}{}_{b}) e^{b}_{\alpha}, \quad \delta g_{\alpha\beta} = \Omega g_{\alpha\beta}.
\end{equation}
We see that $\Lambda^{ab}$ generates infinitesimal Lorentz rotations of the frame fields and $\Omega$ generates infinitesimal local conformal transformations.  Hence, the invariance of the action under (\ref{eq:gauge 3}) means that our model is conformally invariant.

The following gauge transformation generates a conformal transformation of the metric and preserves the gauge condition $q_\alpha = 0$:
\be \epsilon^{A} \mathbf{J}_{A} = -2 e^{a\alpha} \partial_{\alpha} \Omega \, \mathbf{K}_{a} + \Omega \mathbf{D}.
\ee
Under this, the torsion tensor transforms as
\be
\delta T^a_{\alpha\beta} = \tfrac{1}{2} T^a_{\alpha\beta} \Omega.
\ee
So, under transformations that preserve the $q_\alpha = 0$ gauge, the torsion tensor is gauge invariant, and the torsion-free sector is preserved under this restricted group of gauge transformation.  

\section{Explicit equations of motion}

\subsection{With torsion}
In a gauge where $q_{\alpha} = 0$, we can re-write the components of the equation of motion (\ref{eq:EOM 1}) explicitly in terms of spacetime tensors:
\begin{eqnarray}
	\nn a^{\alpha\nu} & = & \hat\nabla_{\mu}T^{\alpha\mu\nu} + R^{\alpha\nu} - \tfrac{1}{2}(f^{(\alpha\nu)}+\tfrac{1}{2} f g^{\alpha\nu}), \label{eq:a eqn} \\
	\nn b^{\alpha\nu} & = & \hat\nabla_{\mu}\theta^{\alpha\mu\nu} -f_{\lambda\mu}\Phi^{\alpha\lambda\mu\nu} - f^{\alpha}{}_{\mu}\f^{\mu\nu} - \tfrac{1}{2} \tau^{\alpha\nu}, \label{eq:b eqn} \\
	\nn c^{\alpha\beta\nu} & = & \hat\nabla_{\mu} \Phi^{\alpha\beta\mu\nu} + \tfrac{1}{2} \theta^{[\alpha\beta]\nu} - \tfrac{1}{2} f^{[\alpha|\mu|} T^{\beta]\nu}{}_{\mu}, \label{eq:c eqn} \\
	\nonumber d^{\nu} & = & (2\hat\nabla_{\mu}+\nabla_{\mu})\f^{\mu\nu} + \tfrac{1}{2}\nabla_{\mu}(f^{(\mu\nu)}-g^{\mu\nu} f) \\ & &  +2 \f_{\mu\sigma}T^{[\sigma\mu]\nu}. \label{eq:EOM 2}
\end{eqnarray}
Here, $R_{\alpha\beta}$ is the Ricci tensor and we have defined
\begin{align}
	\theta^{\mu\alpha\beta} & := \nabla^{\alpha} f^{\mu\beta} - \nabla^{\beta} f^{\mu\alpha} + f^{\mu\sigma}T_{\sigma}{}^{\alpha\beta}, \\
	\Phi^{\alpha\beta\gamma\delta}  & := R^{\alpha\beta\gamma\delta} - \tfrac{1}{2}( g^{\gamma[\alpha} f^{\beta]\delta}  - g^{\delta[\alpha} f^{\beta]\gamma} ).
\end{align}

\subsection{Without torsion}

In general, solutions of the equations of motion (\ref{eq:EOM 2}) will have non-zero torsion.  However, in the reminder of the paper we will concentrate on the torsion-free sector of the solution space.  In future work, we will explore the more general case in detail.

If there is no torsion, equation (\ref{eq:torsion def}) implies that:
\begin{equation}
	\di_{[\beta} e_{\gamma]}^{a} + \omega^{ab}_{[\beta} e_{\gamma]b} = 0.
\end{equation}
Making use of the fact that $\omega^{(ab)}_{\alpha} = 0$, this can be solved to give the connection one-forms in terms of the tetrad and its derivatives:
\begin{equation}
	\omega^{ab}_{\beta} = \tfrac{1}{2} e^{\alpha a} e^{\gamma b} (\xi_{\beta\gamma\alpha} - \xi_{\alpha\beta\gamma} - \xi_{\gamma\alpha\beta}),
\end{equation}
where
\begin{equation}
	\xi_{\alpha\beta\gamma} = e_{\alpha a} \di_{\beta} e_{\gamma}^{a} - e_{\alpha a} \di_{\gamma} e_{\beta}^{a}.
\end{equation}
Making note of $g_{\alpha\beta} = e_{\alpha a} e^{a}_{\beta}$, this can be re-written as
\begin{equation}
	\omega^{ab}_{\beta} = \tfrac{1}{2} e^{\alpha a} e^{\gamma b} ( \di_{\beta} g_{\gamma\alpha} + \di_{\gamma} g_{\alpha\beta} - \di_{\alpha} g_{\beta\gamma} - 2e_{\alpha c} \di_{\beta} e_{\gamma}^{c} ).
\end{equation}
Substituting this into the second member of (\ref{eq:tetrad postulate}), we get
\begin{equation}\label{eq:Levi-Civita}
	\Gamma^{\alpha}{}_{\beta\gamma} = \tfrac{1}{2} g^{\alpha\rho} ( \di_{\beta} g_{\gamma\rho} + \di_{\gamma} g_{\rho\beta} - \di_{\rho} g_{\beta\gamma} ).
\end{equation}
Hence, if the torsion is zero we find that $\Gamma^{\alpha}{}_{\beta\gamma}$ reduces to the Levi-Civita connection and $\nabla_{\alpha}$ is the familiar covariant derivative used in general relativity.  Of course we could have come to this conclusion without any calculations:  We have \emph{a priori} assumed that our connection is metric compatible, so the additional restriction that the torsion vanishes means that we must necessarily recover (\ref{eq:Levi-Civita}).

When we assume that $T_{\alpha\beta\gamma} = 0$ in the equations of motion (\ref{eq:EOM 2}), several simplifications occur.  The first member of (\ref{eq:EOM 2}) can be algebraically solved for $f_{(\alpha\beta)}$:
\begin{equation}
	f_{(\alpha\beta)} = 4S_{\alpha\beta} - 2\bar{a}_{\alpha\beta}.
\end{equation}
where $\bar{a}_{\alpha\beta} := a_{\alpha\beta} - \tfrac{1}{6} g_{\alpha\beta} a$, $a=a^{\mu}{}_{\mu}$, and
\begin{equation}
S_{\alpha\beta} := \tfrac{1}{2}(R_{\alpha\beta} - \tfrac{1}{6} R g_{\alpha\beta})
\end{equation}
is known as the Schouten tensor.  The third and fourth members of (\ref{eq:EOM 2}) yield a consistency condition satisfied by matter fields to ensure $T_{\alpha\beta\gamma} = 0$,
\begin{equation}\label{eq:div a}
	0 = \nabla^{\alpha} a_{\alpha\beta} + c_{\beta\alpha}{}^{\alpha} - \tfrac{1}{2} d_{\beta}.
\end{equation}
as well as relations resembling Maxwell's equations in the presence of electric and \emph{magnetic} charges:
\begin{align}\label{eq:Maxwell 1}
	\nabla^{[\alpha} \f^{\beta\nu]} & = \tfrac{1}{9} g^{\nu\beta} c^{\alpha\mu}{}_{\mu} - \tfrac{1}{9} g^{\nu\alpha} c^{\beta\mu}{}_{\mu} - \tfrac{1}{3} c^{\alpha\beta\nu}, \\
	\nabla^{\alpha} \f_{\alpha\beta} & = \tfrac{1}{2} d_{\beta} - \tfrac{1}{3} c_{\beta\alpha}{}^{\alpha} - \tfrac{1}{6} \nabla_{\beta}a.
\end{align}
Finally, the second member of (\ref{eq:EOM 2}) gives
\begin{equation}\label{eq:Bach EOM}
	B^{\alpha\nu} = -\tfrac{1}{4} b^{\alpha\nu} + \nabla_{\mu}(\nabla^{[\nu}\bar{a}^{\mu]\alpha} + \nabla^{[\nu}\f^{\mu]\alpha}) + \mathscr{Q}^{\alpha\nu},
\end{equation}
where
\begin{equation}\label{eq:Bach def}
B_{\mu\nu} = -\nabla^{\alpha} \nabla_{\alpha} S_{\mu\nu} + \nabla^{\alpha} \nabla_{\mu} S_{\alpha\nu} + C_{\mu\alpha\nu\beta} S^{\alpha\beta},
\end{equation}
is the Bach tensor, $C_{\alpha\beta\gamma\delta}$ is the Weyl tensor, and $\mathscr{Q}^{\alpha\nu}$ is a tensor quadratic in $\f_{\alpha\beta}$, $a_{\alpha\beta}$ and the curvature:
\begin{align}\nn
\mathscr{Q}^{\alpha\nu} = & \tfrac{1}{2} a_{\lambda\mu} C^{\alpha\lambda\mu\nu} - \tfrac{1}{8} \tau^{\alpha\nu}-\tfrac{1}{2}(2S_{\lambda\mu} - \bar{a}_{\lambda\mu} + \f_{\lambda\mu}) \times \\ \nn & (g^{\lambda[\mu} \bar{a}^{\nu]\alpha} - g^{\alpha[\mu} \bar{a}^{\nu]\lambda}  + g^{\alpha[\mu} \f^{\nu]\lambda}  -g^{\lambda[\mu} \f^{\nu]\alpha} \\ & +g^{\alpha\lambda} \f^{\mu\nu}).
\end{align}

\section{Weak fields}  We now consider the linearization of the torsion-free sector of the model about a Minkowski background in which $R_{\alpha\beta\gamma\delta}$, $\f_{\alpha\beta}$ and all matter fields vanish.  For simplicity, we will consider perturbative matter sources with
\begin{equation}
	c_{\alpha\beta\gamma}=0=d_{\alpha} \quad \Rightarrow \quad \di^{\alpha} a_{\alpha\beta} = 0.
\end{equation}
As in \cite{Mannheim:1992tr,*Mannheim:1999bu,*Mannheim:2007ug,*Mannheim:2011ds}, we define a traceless metric perturbation $H_{\alpha\beta}$:
\begin{equation}
	g_{\alpha\beta} = \eta_{\alpha\beta}+h_{\alpha\beta}, \quad H_{\alpha\beta} = h_{\alpha\beta} - \tfrac{1}{4} \eta_{\alpha\beta} h.
\end{equation}
Under these assumptions and expanding to linear order, we find that
\begin{align}\label{eq:Maxwell 3}
	\nabla^{[\alpha} \f^{\beta\nu]} = 0, \quad \nabla^{\alpha} \f_{\alpha\beta} =  - \tfrac{1}{6} \nabla_{\beta}a, \quad \Box a =0,
\end{align}
and
\begin{multline}\label{eq:H wave equation}
	-\tfrac{1}{4} ( \tfrac{2}{3} \di_{\alpha} \di_{\beta} \di^{\mu} \di^{\nu} H_{\mu\nu} - 2\Box \di^{\nu} \di_{(\alpha}H_{\beta)\nu} + \tfrac{1}{3} \eta_{\alpha\beta} \di^{\mu} \di^{\nu} \Box H_{\mu\nu} \\ + \Box^{2} H_{\alpha\beta}) = \tfrac{1}{2} \Box a^\text{TF}_{\alpha\beta} +\tfrac{1}{4} b_{\alpha\beta} + \tfrac{1}{6} \di_{\alpha}\di_{\beta}a.
\end{multline}
Here, $a^\text{TF}_{\alpha\beta}$ indicates the trace-free part of $a_{\alpha\beta}$.  We see that the trace of $a_{\alpha\beta}$ acts as a source for both $\f_{\alpha\beta}$ and $H_{\alpha\beta}$.  We can hence consider $H_{\alpha\beta}$ to be the sum of contributions sourced by $a$ and $a^\text{TF}_{\alpha\beta}$, respectively.  Concentrating on the latter, we can impose the transverse gauge condition $\di^{\alpha} H_{\alpha\beta} = 0$ to simplify the equations of motion.  For static sources and fields, the equations can be solved explicitly via Green's functions in this gauge:
\begin{equation}
H_{\alpha\beta}(\mathbf{r}) = \int d^{3}\mathbf{r'} \frac{ a^\text{TF}_{\alpha\beta} (\mathbf{r'})}{2\pi |\mathbf{r} -\mathbf{r'}| } +\int d^{3}\mathbf{r'} \frac{ |\mathbf{r} -\mathbf{r'}|}{8\pi} b_{\alpha\beta}(\mathbf{r'}).
\end{equation}
We see that the metric perturbations sourced by $a_{\alpha\beta}^\text{TF}$ fall-off as inverse distance, just as in the static weak field limit of general relativity.  However, the perturbations sourced by $b_{\alpha\beta}$ increase proportionally with distance, implying a modification of the gravitational interaction on long wavelengths.  The existence of distinct short and long range gravitational forces is directly related to the appearance of $\Box a^\text{TF}_{\alpha\beta}$ in (\ref{eq:H wave equation}) and not $a^\text{TF}_{\alpha\beta}$, which in turn follows from (\ref{eq:Bach EOM}).

If we do not impose the transverse condition ($\di^{\alpha} H_{\alpha\beta} = 0$) and assume that $a_{\alpha\beta}$ and $b_{\alpha\beta}$ both have $\delta$-function support at the origin, we can solve (\ref{eq:H wave equation}) explicitly:
\begin{gather}
	ds^{2} = -(1+2\phi) dt^{2} + (1-2\psi) (dx^{2}+dy^{2}+dz^{2}), \\ \frac{\phi+\psi}{2} = -\frac{r_{a}}{r} - \frac{r}{r_{b}}.
\end{gather}
Here, $r_{a}$ and $r_{b}$ are constants proportional to the amplitude of the $\delta$-functions in $a_{\alpha\beta}$ and $b_{\alpha\beta}$, respectively.   Note that $r_{a}$ and $r_{b}$ are not necessarily positive; their signs cannot be determined in the absence of a specific matter model.  Also note that the equation of motion (\ref{eq:H wave equation}) does not fix the difference $\phi-\psi$ of the metric potentials.  This is because the gauge transformations (\ref{eq:gauge 2}) gives $\delta h_{\alpha\beta} = \epsilon \eta_{\alpha\beta}$, which implies:
\begin{equation}
	\delta(\phi+\psi) = 0, \quad \delta(\phi-\psi)=\epsilon;
\end{equation}
i.e., $\phi+\psi$ is gauge invariant while $\phi-\psi$ is a purely gauge degree of freedom.  We can use this freedom to impose a parametrized-post-Newtonian (PPN) \cite{Will:2014xja} gauge condition $\phi = -r_{a}/r$ from which it follows that the standard PPN parameter $\gamma$ is
\begin{equation}
	\gamma = \frac{\psi}{\phi} = 1+\frac{2r^{2}}{|r_{a}r_{b}|}.
\end{equation}
Hence, in order to satisfy the Cassini constraint $|\gamma-1| \lesssim 10^{-5}$, we need $|r_{a}r_{b}| \gg r^{2}$ in the solar system.

One may be concerned that when we assume that $a_{\alpha\beta} \propto \delta(\mathbf{r})$, the source term on the righthand side of (\ref{eq:H wave equation}) contains derivatives of a $\delta$-function; i.e., our mechanism for recovering Newton's law involves the same highly singular sources as envisioned by Mannheim et al \cite{Mannheim:1992tr,*Mannheim:1999bu,*Mannheim:2007ug,*Mannheim:2011ds}.  However, there is a subtle but important distinction: Mannheim assumes the matter density for pointlike sources (and hence stress-energy tensors) involves factors of the form $\roarrow{\nabla}^{2} \delta(\mathbf{r})$, which are hard to obtain from the variation of a non-singular classical action.  In our calculations, matter fields obtained from variation of the action (i.e., $a_{\alpha\beta}$, $b_{\alpha\beta}$, etc.) have no worse than $\delta$-function singularities, and hence imply the action is finite.  (Further discussion of the viability of the Newtonian limit in Mannheim's theory, including the possible need for extended sources, can be found in Refs.\ \cite{Flanagan:2006ra,Yoon:2013rxa,*Mannheim:2015gba}.\footnote{It should be noted that the conclusions of \cite{Flanagan:2006ra} have been rebutted in \cite{Mannheim:2007ug}.})

\section{Discussion}

We have reconsidered a gauge theory of gravity in which the gauge group is the conformal group SO(4,2) and the action is of the Yang-Mills form, quadratic in the curvature.  By identifying the ``background metric'' with the fields gauged by the translation generator, the full SO(4,2) gauge invariance is broken to that generated by the Lorentz rotations, special conformal transformations and dilatations.  Under certain restrictions, the vacuum equations of motion of our model are solved by the solutions of the vacuum equations of Weyl squared gravity.

We then considered the linearization about torsion-free Minkowski spacetime.  We found that matter which couples to the generators of special conformal transformations induces gravitational forces which fall off as $r^{-2}$, as in the weak field limit of GR.  Conversely, matter that couples directly to the vierbein induces forces which are asymptotically constant in the far field.  Unlike Weyl's original conformal gravity (in the absence of the highly singular sources proposed in \cite{Mannheim:1992tr,*Mannheim:1999bu,*Mannheim:2007ug,*Mannheim:2011ds}), the theory is potentially consistent with solar systems tests of gravity (with judicious parameter choices).  Furthermore, the long range behaviour of the gravitational interaction could provide an explanation of the late time acceleration of the universe.  The fact that matter with nonstandard coupling is responsible for Newtonian gravity is an exotic feature of our model that deserves further study.  Finally, the theory may be perturbatively renormalizable, since the gauge coupling constant (in four dimensions) is dimensionless.

However, we must state an important caveat:  As (\ref{eq:H wave equation}) shows, the linearized sector of the theory is governed by a fourth-order wave equation.  According to ``standard folklore'' such a theory should be prone to ghost instabilities upon quantization \cite{Stelle:1976gc}; however we note that this is a controversial assertion and there are known counter examples \cite{Bender:2007wu,*Bender:2008gh}.  In future work, we plan to study this issue in detail by examining the canonical form of the theory and its quantization.  Also of interest for further work is the role of torsion and $\f_{\alpha\beta}$ in astrophysics at both the linear and non-linear level, as well as cosmological solutions.

We conclude by noting that local conformal symmetry is clearly not an observed property of low-energy physics.  Therefore, our model will require a mechanism to break this symmetry at low energy in order to be consistent with familiar phenomena.  There are several ways this can be accomplished, one of which involves coupling the model to non-conformally invariant classical matter.  Or, as advocated by Mannheim \cite{Mannheim:2010hk} and t'Hooft \cite{Hooft:2014daa}, one can assume conformal symmetry is broken spontaneously in analogy to the breaking of electroweak symmetry in the standard model via the Brout--Englert--Higgs mechanism.  That is, the vacuum state of some quantum field picks out a preferred conformal frame.  We hope to report on such issues in the future.

\begin{acknowledgments}

This work was supported by NSERC of Canada.

\end{acknowledgments}

\appendix

\section{Reduction to Weyl squared gravity}\label{sec:weyl squared}

In this appendix, we examine our model under the following (somewhat restrictive) assumptions:
\begin{enumerate}[label=(\emph{\roman*}),noitemsep]
	\item Torsion-free: $T_{\alpha\beta\gamma} = 0$;
	\item $q_{\alpha} = 0$;
	\item $\f_{\mu\nu} = 0$; and
	\item $S_\text{m}$ only depends on $g_{\alpha\beta}$ and matter fields, and none of the other components of $\mathbf{A}_{\alpha}$; i.e, $a_{\alpha\beta}=c_{\alpha\beta\gamma}=d_{\alpha}=0$.
\end{enumerate}
We will see that these lead us to Weyl-squared gravity.

Under these circumstances, the action (\ref{eq:spacetime action}) reduces to
\begin{equation}\label{eq:torsion free action}
	S = \frac{2}{g_\text{YM}^{2}} \int d^{4}x \sqrt{-g} (R_{\alpha\beta\gamma\delta} - \tfrac{1}{2} \phi_{\alpha\beta\gamma\delta})^{2} + S_\text{m}.
\end{equation}
Referring to the definition (\ref{eq:phi def}) of $\phi_{\alpha\beta\gamma\delta}$, we see that this action contains no derivatives of $f_{\alpha\beta}$.  Hence, $f_{\alpha\beta}$ is essentially a Lagrange multiplier.  Variation of the action with respect to $f_{\alpha\beta}$ yields:
\begin{multline}
	\frac{\delta S}{\delta f^{\alpha\beta}} \delta f^{\alpha\beta} = \frac{8}{g_\text{YM}^{2}} \int d^{4}x \sqrt{-g} \\ \times [ R_{\alpha\beta} - \tfrac{1}{2} \left(f_{\alpha\beta} + \tfrac{1}{2} f g_{\alpha\beta} \right) ] \delta f^{\alpha\beta}.
\end{multline}
Setting this equation to zero yields a constraint, which after some algebra gives:
\begin{equation}
	f_{\alpha\beta} = 2 \left( R_{\alpha\beta} - \tfrac{1}{6} R g_{\alpha\beta} \right).
\end{equation}
Substituting this constraint back into the action (\ref{eq:torsion free action}) gives
\begin{equation}\label{eq:torsion free action 2}
	S = \frac{2}{g_\text{YM}^{2}} \int d^{4}x \sqrt{-g} \, C_{\alpha\beta\gamma\delta}  C^{\alpha\beta\gamma\delta} + S_\text{m},
\end{equation}
where $C_{\alpha\beta\gamma\delta}$ is the ordinary Weyl-tensor.  Now, variation with respect to the metric yields
\begin{equation}
	\delta S = \frac{4}{g_\text{YM}^{2}}  \int d^{4}x \sqrt{-g} \, \left(-B^{\alpha\beta} - \tfrac{1}{4} g_\text{YM}^{2} \mathscr{T}^{\alpha\beta}\right)\delta g_{\alpha\beta},
\end{equation}
where $\mathscr{T}^{\alpha\beta}$ is the ordinary stress energy tensor, and the Bach tensor is defined in (\ref{eq:Bach def}).  We then obtain the field equation
\begin{equation}\label{eq:conformal EOM}
	B_{\mu\nu} = -\tfrac{1}{4} g_\text{YM}^{2} \mathscr{T}_{\mu\nu}.
\end{equation}
This is the equation originally studied by Bach \cite{bach}.  Notice that the Bach tensor is traceless, which places restrictions on the form of material sources of the form $g^{\alpha\beta} \mathscr{T}_{\alpha\beta} = 0$.

\vfill

\bibliography{gravity_yang_mills}

\vfill

\end{document}